# Topological charge analysis of dynamic process of transition to Néel-type skyrmion: role of domain wall skyrmions

Soong-Geun Je[*]

*Department of Physics, Chonnam National University, Gwangju 61186, Republic of Korea*

*e-mail: sg.je@jnu.ac.kr

**Abstract**

Magnetic skyrmions are intriguing topological spin textures that promise future high-density spintronic devices. The creation of magnetic skyrmions has been understood based on the energetics of skyrmions, but the detailed dynamic process of the skyrmion creation remains unclear. Here topological evolution in conversion from uneven domains to Néel-skyrmions is investigated using micromagnetic simulations. We find that, rather than the overall topological charge, annihilation of novel topological defects, i.e., recently suggested domain wall skyrmions, dominantly govern the skyrmion creation process. Also, the topological charge evolution is interpreted in terms of the number and the combination of such topological defects.

# 1. Introduction

Magnetic skyrmions are magnetic pseudo-particles that have whirling spin configurations possessing a topological charge $Q$ of ±1 [1]. Owing to their nanoscale size [2], durability by a topological barrier [3], and electrical controllability [4,5], skyrmions are currently attracting considerable attention as robust information carriers for future spintronic storage and computing applications [4,6,7].

To realize such applications, establishing skyrmion creation schemes and understanding the underlying creation mechanisms are crucial. Skyrmions have been generated by commonly injecting current pulses, which induce heating or spin torques [8-11], or by shining a laser beam to exploit localized heating [12-14]. Among these writing schemes, the thermal creation of skyrmions can be explained based on the framework of the energetics of skyrmions where the thermal effect allows the system to reach the stable skyrmion state by overcoming energy barriers [15,16]. The energetics of skyrmions provides an intuitive understanding of the thermal transition from a ferromagnetic state to the skyrmion state. However, it does not describe the detailed dynamic process during the transition.

Meanwhile, in Ref. [14], the dynamic process of the skyrmion creation was presented by micromagnetic simulations. In the simulations, the skyrmion state was obtained by relaxing the system from a random magnetization state. It was shown that, at the very beginning of the relaxation, plenty of reversed domains are created, but not all the reversed domains transform into skyrmions. This observation raises questions under what conditions and how the domains evolve into skyrmions, and why most of them disappear.

A variety of spin textures can be classified based on their topological charges, which count the number of times the spin configuration wraps around a unit sphere. Since magnetic skyrmions are characterized by a topological charge $Q$ of ±1, the evolution from a domain possessing other topological charges to skyrmions should accompany a change in the topological charge mediated by the emergence and annihilation of topological defects [1,17,18]. Thus, it is interesting to investigate the evolution of topology and the associated spin structures as well as involved topological defects to reveal the dynamic process of skyrmion generation [19].

In this letter, we study the evolution of topological charge and associated spin structures during the dynamical skyrmion creation, where the skyrmion state is reached by relaxing a random magnetization state. Micromagnetic simulations and topological charge analysis show that rather than the topological charge is conserved during the process, the overall topological charge changes with discrete jumps of a unit topological charge. We found that the unit jumps result from the annihilation of a novel type of topological defects in the domain walls, i.e., recently proposed domain wall skyrmions [20,21] in a system with a finite Dzyaloshinskii-Moriya interaction (DMI). Also, the number and combination of such topological defects are found to govern the skyrmion creation process. This work provides insight into the process of dynamically creating skyrmions, highlighting the role of domain wall skyrmions.

## 2. Simulation

For this study, we carry out micromagnetic simulations using the Object Oriented MicroMagnetic Framework (OOMMF) public code with DMI package [22,23]. Simulation parameters are chosen as: saturation magnetization $M_s = 1.0 \times 10^6$ A/m, exchange stiffness $A_{ex} = 1.2 \times 10^{-11}$ J/m, uniaxial perpendicular anisotropy $K_u = 6.8 \times 10^5$ J/m$^3$, and damping constant $\alpha = 0.5$. Particularly, a finite DMI constant $D = 1.0$ mJ/m$^2$ is introduced to study Néel-type skyrmions which have technological merits [4]. The simulation region is $0.4 \times 0.4$ μm$^2$ and the cell size is $2 \times 2 \times 0.3$ nm$^3$. To mimic the skyrmion creation from a thermally demagnetized region [14], a random magnetization state is embedded in the center of the simulation area and taken as an initial state. The system is then relaxed in the presence of an out-of-plane magnetic field of +8 mT.

## 3. Results and discussions

Fig. 1(a) displays a typical example of the relaxation process, and Fig. 1(b) shows the total topological charge of the entire area with respect to the simulation time. Here the topological charge $Q = 1/4\pi \int m \cdot (\partial_x m \times \partial_y m) dxdy$ is employed [1]. After initial fluctuations, three isolated domains appear (0.638 ns), but only one domain finally evolves into a single skyrmion with its topological charge contribution -1 to total $Q$.

As the total $Q$ gets saturated, one can see a single rise and fall of the total $Q$, as indicated by the yellow, green, and magenta solid circles in Fig. 1(b). Looking at the corresponding magnetization images [designated by colored circles in Fig. 1(a)], one can notice that the unit jumps of ±1 are closely related to the domain morphology, specifically the annihilation of the kinks [yellow and green arrows in Fig. 1(a)].

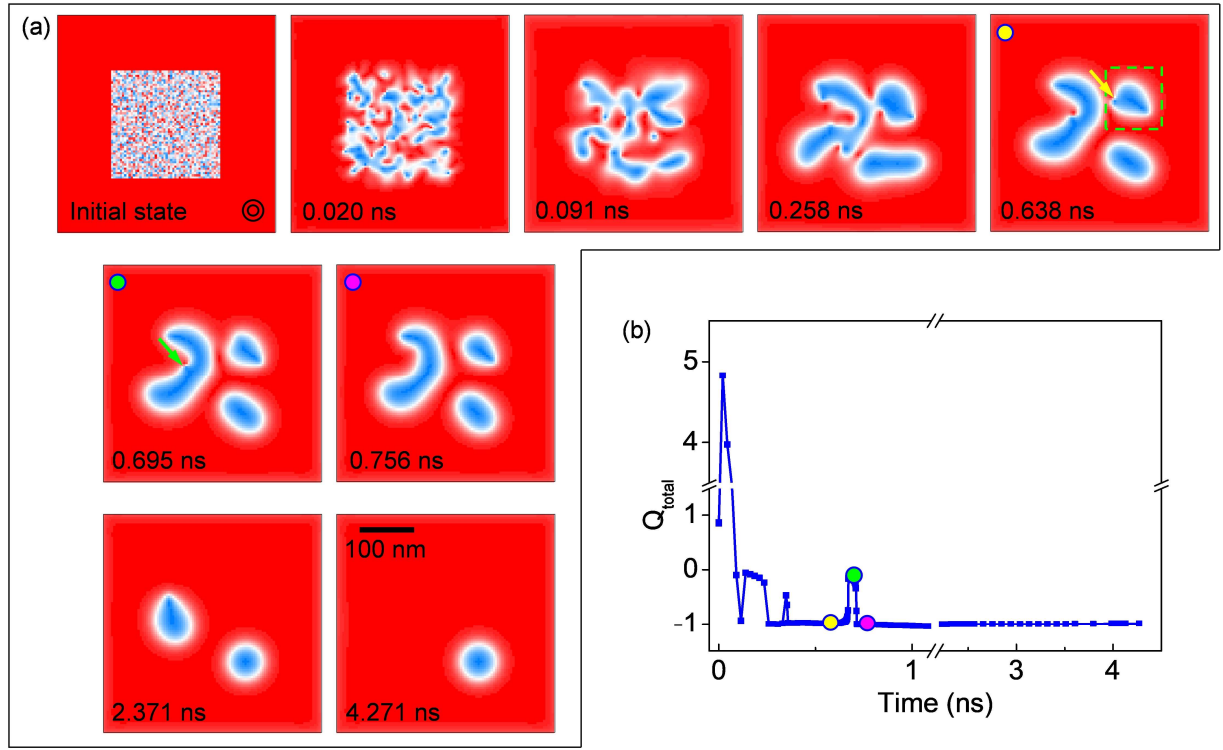

**Figure 1.** (a) A typical example of the skyrmion generation by relaxation from a random-magnetization state. Red (blue) indicates magnetization pointing up (down). (b) Total topological charge of the simulation in Fig. 1(a) with respect to the simulation time. The colored circles refer to the corresponding simulation images in Fig. 1(a).

To more investigate the relationship between the topological charge $Q$ and the detailed spin texture of the kink, we first examine the evolution of the isolated single domain marked by the green square in Fig. 1(a). Fig. 2(a) shows the topological charge of the domain as a function of time. Initially, the topological charge $Q$ is -1, which is equivalent to that of the skyrmion. However, in turn, the topological charge $Q$ becomes zero, and the domain disappears in the end as shown in Fig. 1(a). This is interesting because the topological charge $Q = -1$ itself does not guarantee the evolution of the domain into the skyrmion having the same topological charge.

Figs. 2(b) and 2(c) [Figs. 2(d) and 2(e)] respectively display the domain morphology and topological charge density of the domain with $Q = -1$ ($Q = 0$). Note that the kink in Fig. 2(b) exhibits the intense negative topological charge density [dark blue in Fig. 2(c)]. The annihilation of the kink results in the topological charge transition from -1 to 0, indicating that the kink possesses the topological charge $Q = -1$. The detailed spin texture of the kink is shown in Fig. 2(f) and the spin texture after the annihilation is given in Fig. 2(g) for comparison. In Fig. 2(f), the magnetization is rapidly reversed at the kink, opposing the DMI-favored Néel wall magnetization outside the kink.

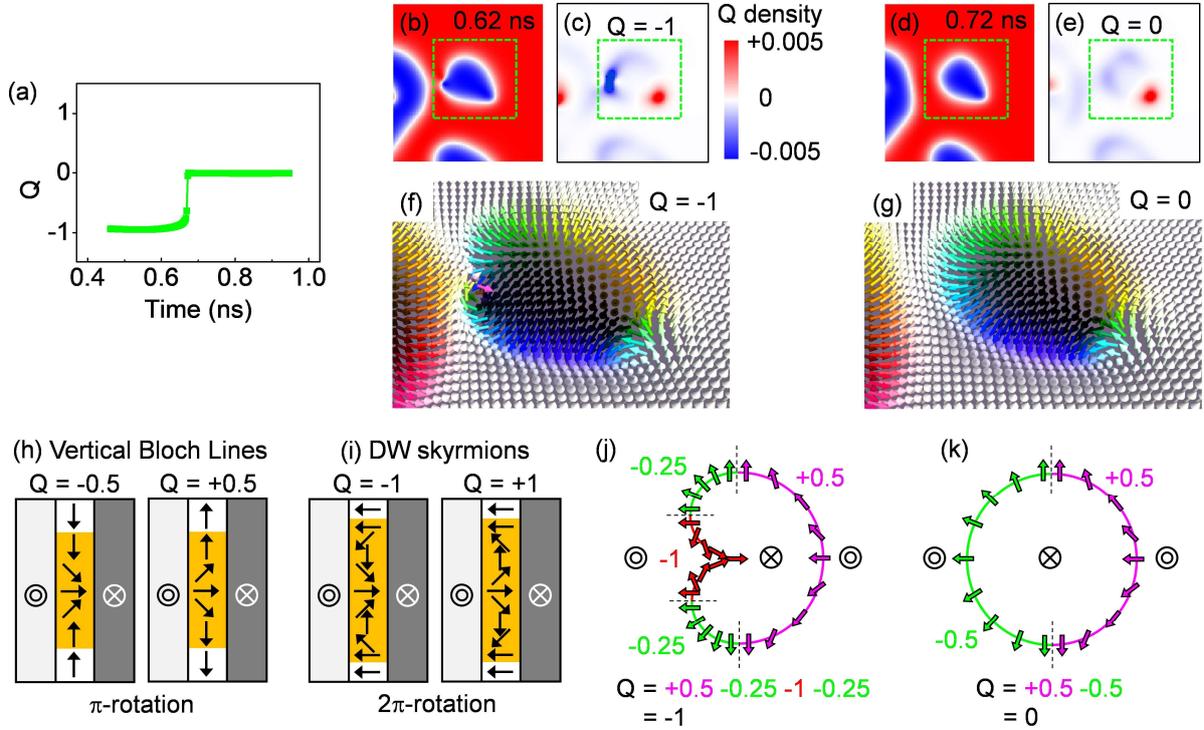

**Figure 2.** (a) The evolution of the topological charge in the area marked by the green square in Fig. 1(a). (b), (c) The domain morphology and the topological charge density of a domain at $Q = -1$ before the jump in Fig. 2(a). (d), (e) The domain morphology and the topological charge density of a domain at $Q = 0$ after the jump in Fig. 2(a). (f), (g) Detailed spin structures of Figs. 2(b) and 2(d), respectively. Muview2 code is used for visualization [24]. (h), (i) Schematics depicting the structure of vertical Bloch lines (h) and DW skyrmions (i). (j), (k) Schematics of the domain structure in Figs. 2(f) and 2(g), respectively.

We note that the spin structure of the kink corresponds to the domain wall (DW) skyrmions, which are recently proposed a different type of magnetic topological defects that reside in DW [20,21]. In the absence of DMI, a Bloch-type wall is favored and vertical Bloch lines with π-rotation [Fig. 2(h)] are commonly observed topological defects in the DW [25]. However, a sufficiently strong DMI stabilizes Néel-type DWs with fixed chirality [26]; thus, allowed magnetic excitations in the DW are 2π-rotation along the DW as schematically shown in Fig. 2(i). This type of magnetic excitations is the so-called DW skyrmions [20,21]. While the topological charge of vertical Bloch lines is ±0.5, that of the DW skyrmions is ±1 depending on rotation direction. The DW skyrmion might be seen as a pair of vertical Bloch lines, however, we note that individual annihilation of vertical Bloch lines during the DW skyrmion annihilation is not observed, meaning the interpretation based on the DW skyrmion is more reasonable in this system with strong DMI.

Noting that the DW skyrmions are involved in the topological charge evolution, we now connect the topological charge $Q$ to the detailed spin structures. Fig. 2(j) [Fig. 2(k)] is the schematic of Fig. 2(f) [Fig. 2(g)]. The spin structure in Fig. 2(f) consists of the vertical Bloch line of $Q = +0.5$, two quadrants of a Néel-type skyrmion (-0.25 each), and the DW skyrmion of $Q = -1$, thus resulting in the total $Q = -1$. On the other hand, the DW skyrmion is missing in Fig. 2(k), so the topological charge $Q$ is zero, which is equivalent to the uniform magnetization state. Based on the topological argument, the domain structure in Figs. 2(f) and 2(j) can smoothly evolve into a skyrmion structure. However, the metastable DW skyrmion suddenly annihilates alone so the topological charge of the domain becomes zero, and the domain finally disappears. The above analysis suggests that the DW skyrmion plays an important role in forming the skyrmion and the topological charge evolution.

Next, we present another case that the DW skyrmion determines the final skyrmion state. Fig. 3(a) shows the change of the topological charge of a domain shown representatively in Figs. 3(b)-3(f) along with the detailed spin structures [Figs. 3(g)-3(k)]. The colored circles match the simulation times in Fig. 3(a) to corresponding simulation images. The topological charge of Figs. 3(b) and 3(c) is calculated as -1. Interpreting the corresponding spin configuration in Figs. 3(g) and 3(h) also yields to the same net topological charge ($Q = -1$) because the spin configuration is the combination of the DW skyrmion with $Q = -1$ [green arrow in Fig. 3(b)], anti-skyrmion with $Q = +1$ [yellow arrow in Fig. 3(b)], and skyrmion with $Q = -1$.

As the anti-skyrmion annihilates [Figs. 3(d) and 3(i)], the topological charge of the domain then becomes -2, as schematically depicted in Fig. 3(l). Since the topological charge of -2 is equivalent to that of two separated skyrmions, one might expect the domain turns into two skyrmions. However, different from the expectation based on the topological view, the final state is a single skyrmion [Fig. 3(f)]. Indeed, the spin configuration can transform into two skyrmions if the DW skyrmion (marked by A) merges with the other side of the DW [point B in Fig. 3(l)], thus conserving the topological charge (this case will be subsequently discussed). However, the DW skyrmion annihilates alone without merging, hence the spin configuration becomes the single skyrmion together with the topological charge transition from -2 to -1 as shown in Figs. 3(l) and 3(m).

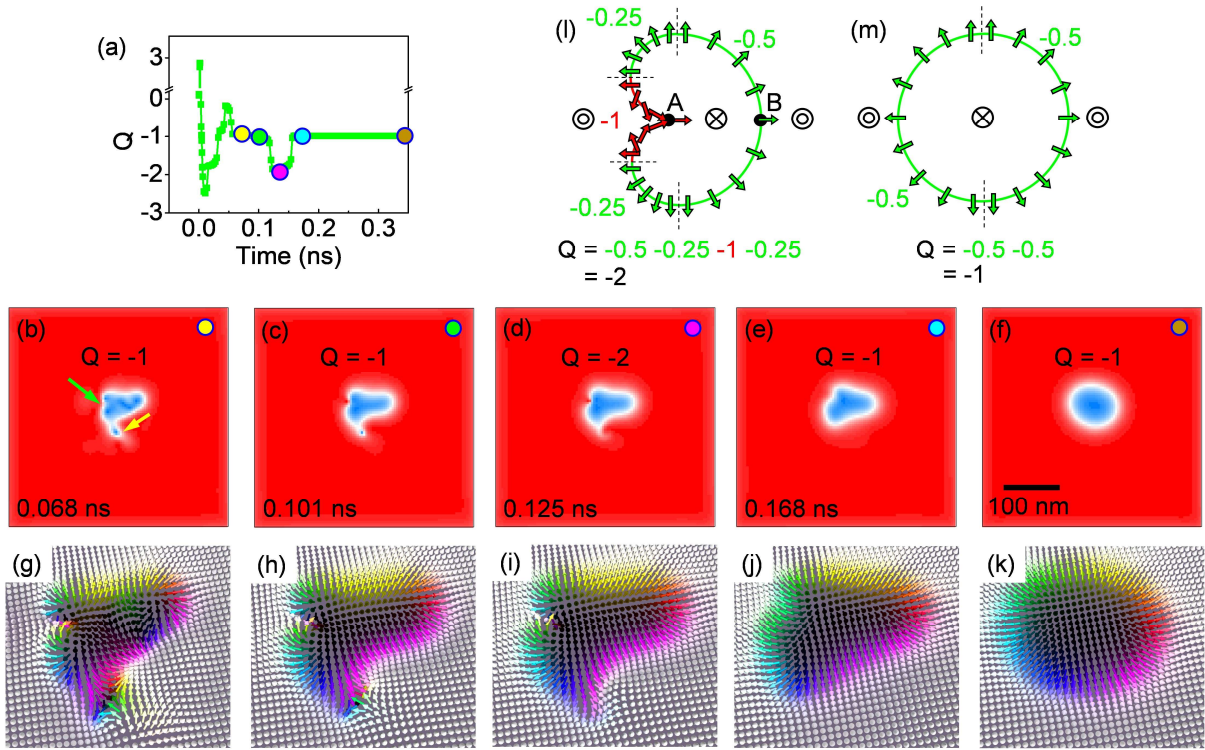

**Figure 3.** (a) Topological charge of the entire area of the simulation representatively given in Figs. 3(b)-3(f). (b)-(f) Evolution of magnetic morphology and the topological charge. The colored circles indicate the simulation time in Fig. 3(a). (g)-(k) The spin configurations of the domains in Figs. 3(b)-3(f). (l), (m) Schematics of the domain structure in Figs. 3(i) and 3(j), respectively.

Finally, we discuss the case where a complex domain with $Q = -1$ splits into two skyrmions rather than evolving into one skyrmion. We track the topological charge of the domain enclosed by the green dashed polygon, as shown in Fig. 4(a). Figs. 4(b)-4(e) show the typical spin configurations and schematics at the several simulation times designated by the colored circles. The complex domain in Fig. 4(b) can be understood in a way that two opposite DW skyrmions are embedded in a skyrmion as the schematic shows. In Fig. 4(c), as the DW skyrmion with $Q = -1$ merges with a DW segment on the opposite side, the domain starts to split into two, but the total topological charge is still -1 because there is a DW skyrmion with $Q = +1$ together with two skyrmions ($Q = -2$). Eventually, the remaining metastable DW skyrmion with $Q = +1$ disappear, producing two stable skyrmions ($Q = -2$) as shown in Figs. 4(d) and 4(e). Through the above process, one can deduce that the sequence of the annihilation of the DW skyrmions and their combination result in different final states, indicating that the DW skyrmions govern the skyrmion creation process.

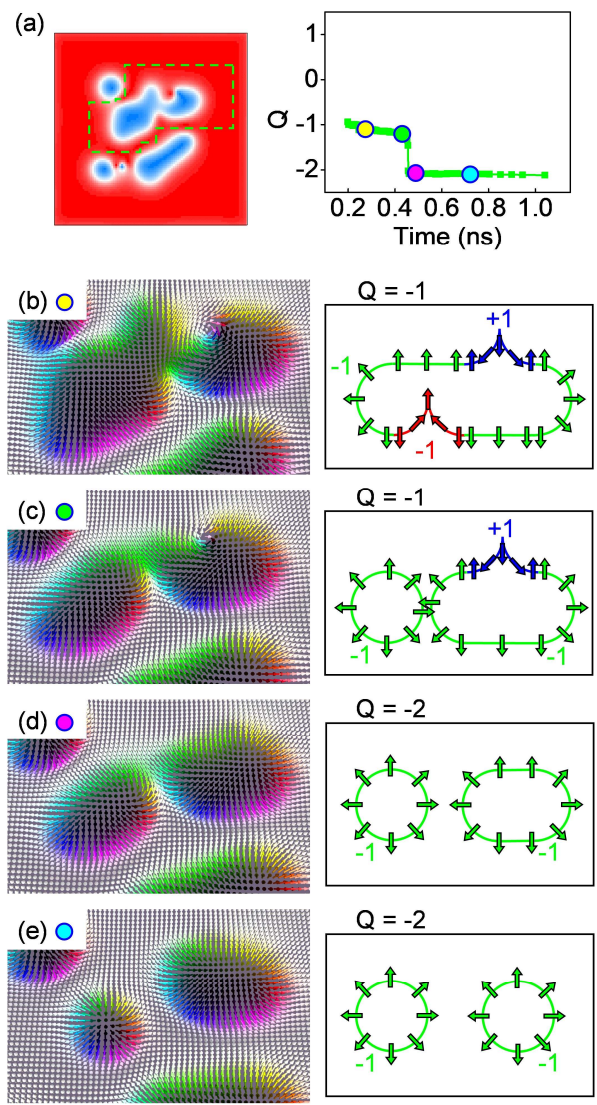

Figure 4. (a) The morphology of the domain of interest enclosed by the green polygon and its topological charge of the area. (b)-(e) The spin structures and their schematics at the points indicated by the colored circles in Fig. 4(a).

## 4. Conclusion

In summary, we investigated the evolution of the topological charge and the detailed spin configuration in the skyrmion generation process using micromagnetic simulations. We found that, in the presence of the DMI, the DW skyrmions emerge and play governing roles in the skyrmion generation process as well as the evolution of the topological charge. Through several typical examples, we could correlate the evolution of the topological charge and detailed spin structures of complex domains, anatomizing the skyrmion creation process. Our work will provide intuitive understanding of the topology of spin textures and their evolution.